\documentclass[preprint,eqsecnum,showpacs,nofootinbib,aps,amsmath,amssymb,tightenlines]{revtex4}

\usepackage{amssymb}
\usepackage[dvips]{graphicx}
\usepackage{color}
\usepackage{times}
\usepackage{bm}         

\newcommand{\mur}[2]{\mu_{#1}\! \left( #2 \right)}

\def\p{\partial}

\def\rsc{{^2\mathcal{R}}}
\def\rscHat{{^2 \hat{\mathcal{R}}}}
\def\ImPsiTwo{{\mathrm{Im}\Psi_2}}
\def\ImPsiTwoHat{{\mathrm{Im}\hat{\Psi}_2}}

\def\lmax{{l_{\rm max}}}
\def\lmaxI{{l^I_{\rm max}}}
\def\lmaxL{{l^L_{\rm max}}}
\def\nmax{{n_{\rm max}}}
\def\nmaxI{{n^I_{\rm max}}}
\def\nmaxL{{n^L_{\rm max}}}
\def\Nlm{\mathcal{N}}
\def\pp{\textrm{\tiny $\parallel$}}

\begin{document}

\title{New method to compute quasi-local spin and other invariants on marginally trapped surfaces}

\author{Michael Jasiulek}
\email{michael.jasiulek@aei.mpg.de}

\affiliation{Max-Planck-Institut f\"ur Gravitationsphysik, Albert-Einstein-Institut, Golm, Germany}

\date{\today}

\begin{abstract}
We accurately compute the scalar 2-curvature, the Weyl scalars, associated quasi-local spin, mass and higher multipole moments on marginally trapped surfaces in numerical 3+1 simulations.
To determine the quasi-local quantities we introduce a new method which requires a set of invariant surface integrals, allowing for surface grids of a few hundred points only.
The new technique circumvents solving the Killing equation and is also an alternative to approximate Killing vector fields.
%
We apply the method to a perturbed non-axisymmetric black hole ringing down to Kerr and compare the quasi-local spin with other methods that use Killing vector fields, coordinate vector fields, quasinormal ringing and properties of the Kerr metric on the surface. Interesting is the agreement with the spin of approximate Killing vector fields during the phase of perturbed axisymmetry.
Additionally, we introduce a new coordinate transformation, adapting spherical coordinates to any two points on the sphere like the two minima of the scalar 2-curvature on axisymmetric trapped surfaces.

\end{abstract}

\preprint{AEI-2009-055}
\maketitle

\section{Introduction}
\label{sec:introduction}

Numerical relativity has undergone a rapid development in the past few years. After the breakthrough of \cite{Pretorius:2005gq,Campanelli:2005dd, Baker:2006yw}, stable longterm simulations of binary black hole (BBH) systems are common practice, besides waveform modelling, to study the close-to-merger spin precession \cite{Campanelli:2006fy, Campanelli:2008nk} or to model the final spin \cite{Rezzolla:2007xa,Rezzolla:2007rd,Barausse:2009uz,Tichy:2008du, Boyle:2007ru} of BBH inspirals \cite{Pollney:2007ss,Baker:2008mj,Herrmann:2007ex,Marronetti:2007wz, Campanelli:2006fg, Scheel:2008rj}.
Recently extensive investigations have been done concerning the formation process
and spin evolution of black holes with accretion disks
\cite{Kiuchi:2009jt, Shibata:2006nm} appearing in fully relativistic
simulations of binary neutron stars \cite{Shibata:1999wi,lrr-2008-7,Baiotti:2008ra},
mixed binaries \cite{Shibata:2006ks,Etienne:2007jg,Duez:2008rb}, rotating neutron star collapse \cite{Shibata:2003agr,Shibata:2000scr,Baiotti:2004wn,Baiotti:2006wm} and rotating supermassiv star collapse \cite{Saijo:2002qt,Shibata:2002br,Saijo:2009ub}. 

In these cases accurate numerical techniques to extract the spin of a BH in a gauge invariant manner are required.
It is common to obtain a rough approximation of the spin through the quasinormal mode oscillation extracted from the gravitational waveform after merger within black hole perturbation theory. 
Another approximation scheme is to integrate the radiated angular momentum contained in the gravitational radiation at `large' coordinate spheres to draw conclusions about the remaining spin of the system given the initial data.

Other methods, as discussed in this paper, use the gauge invariant notation of an apparent horizon (AH) or in more general terms a marginally outer trapped surface (MOTS) which can be located on the spatial slices of the simulation.
There gauge invariant spin and mass can be defined, if an axial Killing vector field (KVF) $\Phi^a$ is present, as in the case of Kerr.
But opposed to the stationary case, the spacetime outside the horizon can be dynamical without spoiling the gauge invariance of these quantities \cite{Ashtekar:2000hw,Ashtekar-Beetle-Lewandowski:2001,Ashtekar:2004cn,Ashtekar-Krishnan:2002}.
The invariant quasi-local spin $J[\Phi^j]$ is given by the surface integral (Brown-York form)
\begin{equation} \label{eq:phi}
 J[\Phi^j] := -\frac{1}{8 \pi} \oint_{S} \Phi^j K_{ij} s^i dA \, ,
\end{equation}
where $dA$ is the 2D area element, $K_{ij}$ the extrinsic curvature of the Cauchy
slice and $s^i$ is the outward-pointing surface normal
on the MOTS denoted by $S$.
In order to obtain $\Phi^j$ the 2D Killing equation has to be solved; if the axisymmetry is perturbed approximate KVFs (aKVFs) have to be computed \cite{Matzner:2003ass,Dreyer:2002mx,Cook:2007akv}, for applications in BBH simulations see \cite{Pollney:2007ss,Scheel:2008rj}. Sometimes, due to computational reasons, the effort of finding a KVF or aKVF is not done and coordinate vector fields are instead used to estimate $J[\Phi^j] \approx J[\Phi^j_{\rm cv}]$, see e.g. \cite{Campanelli:2006fy,Caudill:2006coa}.
Another common set of methods to determine the spin uses properties of the Kerr solution
at the horizon, such as the proper length of the `equatorial' circumference \cite{Smarr:1973sgc} or the extrema of the scalar 2-curvature \cite{Scheel:2008rj}. 

In this paper we present a new, comparatively easy to implement algorithm, which is based on a multipole decomposition of the \emph{rotational} Weyl scalar $\ImPsiTwo$ \cite{Ashtekar:2004gp} in the framework of the isolated and dynamical horizon formalism \cite{Ashtekar-Beetle-Lewandowski:2001, Ashtekar:2004cn, Ashtekar-Krishnan:2002}; for reviews see e.g. \cite{Ashtekar:2004cn, Booth:2005qc, Gourgoulhon:2005ng}.
The dipole term reads
\begin{equation} \label{eq:j1}
 J_1 = - \sqrt{\frac{1}{12\pi}} \frac{A}{4\pi} \oint_S \ImPsiTwo\, Y^{10} (\chi)  \, dA \, ,
\end{equation}
%
where $A$ is the horizon area, $(\chi,\phi)$ an invariant coordinate system \cite{Ashtekar:2004gp} `tied' to the axisymmetry, such that $J_1$ and $J[\Phi^j]$ are identical, and $Y^{10}(\chi)$ is the spherical harmonic $l=1,\, m=0$.
We circumvent the use of invariant coordinates/KVFs and instead use the surface averages $\mu_n$ \footnotemark of the scalar 2-curvature $\rsc$ and $\ImPsiTwo$ to obtain $J_1$ and higher multipole moments
\footnotetext{In statistics $\mu_n$ is called the $n$th \emph{central moment} of the probability distribution of a random variable.}
\begin{equation} \label{eq:sigma}
 \mur{n}{\bullet} := \left< \left( \left< \bullet \right> - \bullet \right)^n \right>, \quad 
 \left< \bullet \right> := \frac{1}{A} \oint_S \bullet\, dA \, ,
\end{equation}
which are well defined, even if the axisymmetry is perturbed and that allow us to benefit from exact numerical integration in order to reduce grid size and numerical error significantly.
The invariant surface integrals $\mur{n}{\rsc},\, \mur{n}{\ImPsiTwo}$ are related to the horizon spin, mass and higher multipole moments by algebraic systems of equations. 
In principal, the $\mu_n$ allow to generalize the horizon multipole moments through solutions of these systems in the absence of axisymmetry.


In order to minimize the numerical error of $\mur{n}{\rsc},\, \mur{n}{\ImPsiTwo}$ accurate numerical computations of the curvature components $\rsc$, $\ImPsiTwo$ and the surface triad \footnotemark on the horizon are required. The horizon is usually given
by $h(\theta,\phi) = \sqrt{\delta_{ij} X^i X^j}$, the Cartesian shape function, where $X^j$ are the Cartesian coordinates at the 2-surface centered at a point inside.
\footnotetext{Note that a `coordinate-induced' surface triad on `large' coordinate spheres (as for wave extraction via $\Psi_4$)
can be easily computed analytically. On the other hand, the coordinate representation of the horizon is a deformed 2-sphere and the computation of derivatives delicate.}
Instead of finite differencing we expand the shape function in terms of a tensor basis to determine Cartesian derivatives off the surface, as commonly used in horizon finding algorithms \cite{Baumgarte:1996iah}.
But opposed to \cite{Baumgarte:1996iah}, we use another basis, which is easier to implement, and exact numerical integration to determine the multipole coefficients of $h(\theta,\phi)$, where \cite{Baumgarte:1996iah} use minimization.

%
%

We apply the new method (in comparison with others) to the dynamical AH of a non-axisymmetric BH \footnotemark ringing down to Kerr in a 3+1 simulation, where we follow the evolution of spin and mass multipoles until their final Kerr values are reached.
%
\footnotetext{We evolve \emph{two puncture} initial data with an initially non-axisymmetric common horizon.}
%

This paper is organized in the following way. In section \ref{sec:kvf} we briefly explain the numerical methods we use to compute KVFs and aKVFs on AHs.
In section \ref{sec:Kerr} we deduce formulas from the Kerr metric to determine Kerr spin and mass from the area and the `equatorial' circumference or the extrema of the scalar 2-curvature on the horizon and give a new formula which requires the surface average $\mu_2(\rsc)$ and that we also apply to our simulations.
In section \ref{sec:multipoles} we show how to use the whole set of $\mu_n$ to extract the multipole spectrum of an axisymmetric isolated horizon.
%
%
In section \ref{sec:computinggeometry} we show how to compute the curvature components $\rsc$, $\Psi_n$ and the surface triad accurately.
In section \ref{sec:num} we explain the setup and initial data of our 3+1 simulation. During the evolution we follow spin, mass and higher multipole moments, compare different methods to measure the spin and test their convergence. 
Notation: Indices $i,j,k$ indicate 3D Cartesian components, indices $a,b,c$ label 2D components on the local horizon grid, letters $l,m$ label spherical harmonics. We indicate dimensionless quantities (mass dimension) with a hat, e.g. $\hat{a}=a/m$, $\rscHat= \rsc \cdot A/(8\pi)$, $\ImPsiTwoHat = \ImPsiTwo \cdot A/(4\pi)$.
\section{Solving the 2D Killing Equation Numerically}%
\label{sec:kvf}%
%
The IH multipole moments are defined in an invariant coordinate system \cite{Ashtekar:2004gp} which requires the knowledge of the axial KVF on the horizon. Our approach does not explicitly require the KVF to extract the IH multipole moments and circumvents the invariant coordinates by using the surface averages $\mu_n(\rscHat),\, \mu_n(\ImPsiTwoHat)$ which can be easily computed in any coordinate system. Nevertheless, in the numerical simulation of section \ref{sec:num} we want to compare our method and hence require the KVF. Therefore, we will briefly explain the techniques we use to solve/approximate the Killing equation.

The induced 2-metric $q_{ab}$ of an spheroid $S$ embedded into Euclidean space admits one rotational Killing vector field $\Phi^a$ which is a solution of the \emph{Killing equation}
%
\begin{equation} \label{eq:killing}
 \mathcal{L}_{\Phi} q_{ab} = 2\, {^2\!D}_{(a} \Phi_{b)}  = 0 \, ,
 \end{equation}
where $^2\!D$ is the induced covariant derivative on $S$. The vector field $\Phi^a$ is unique up to a constant. For Kerr $\Phi^a=\p_\phi$, where $\phi$ is the Boyer-Lindquist coordinate, this constant is fixed such that integral curves have affine length of $2\pi$, thus $\phi \in [0;2\pi]$.
\subsection{Killing Transport Method}
\label{subsec:Ktransport}
In order to solve the Killing equation we apply the \emph{Killing Transport method} \cite{Dreyer:2002mx}, appendix of \cite{wald1984gr}, which is explain in this subsection. 

The method can be roughly divided into three steps: 1. determine a single vector of the KVF at a point on an arbitrary loop on $S$, 2. spread this vector throughout the whole surface, 3. normalize the whole KVF by normalizing an arbitrary integral curve to have affine length of $2\pi$.
The first two steps require the \emph{Killing transport equation}
\begin{eqnarray} \label{eq:transport}
 c^a\, {^2}\!D_a \Phi_b &=& c^a L\, ^2\!\varepsilon_{ab} \\
 c^a\, {^2}\!D_a ( L\, ^2\!\varepsilon_{bc} ) &=&  c^a {^2}\! {R^d}_{cba} \Phi_d \, ,\notag
\end{eqnarray} 
where $^2\!\varepsilon_{ab}$ denotes the Levi-Cevita tensor and $^2\!{R^d}_{cba}$ the 2D Riemann tensor.
The first equation holds, since $ {^2}\!D_{(a}\Phi_{b)} = 0$ if $\Phi_b$ is a KVF and since any two-form on $S$ can be expressed as $L ^2\!\varepsilon_{ab}$, where $L$ is a function. The second equation follows from the first, see \cite{wald1984gr} for details.
Therefore (\ref{eq:transport}) hold for a KVF $\Phi^a$ and the corresponding function $L$ for any vector field $c^a$.

On the other hand, assume that $\Phi_b$ and $L$ were unknown, pick a loop, e.g. the equator $c_{\rm e}\,$, $(\theta=\pi/2,\phi)$ of a spherical coordinate system, pick a point, e.g. $P$, $(\theta=\pi/2,\phi=0)$ and identify $c^a := \p_\phi$ \footnotemark, then (\ref{eq:transport}) becomes an ODE for the unknown $(\Phi_1(\phi), \Phi_2(\phi), L(\phi))$ along $c_{\rm e}$. This defines a linear operator for 3-vectors at $P$. If we pick three arbitrary, linear independent initial vectors at $P$, transport (\ref{eq:transport}) them along the loop to $P$, we obtain a $3\times3$ matrix presentation of this operator. Two components of its eigenvector are the KVF at $P$ (1. step), the third is the auxiliary function $L$ at $P$.
\footnotetext{The resulting KVF is independent of the initial loop, initial point and curve parameter.}
At next this 3-vector is transported with (\ref{eq:transport}) along coordinate lines all over $S$, setting $c^a=\p_\phi$ or $c^a=\p_\theta$ respectively (2. step). Where the transportation equation (\ref{eq:transport}) by construction `conserves' the Killing property. The last step is to normalize the KVF (3. step), where we have to solve the ODE $\p_t \theta = \Phi^1(\theta,\phi),\, \p_t \phi = \Phi^2(\theta,\phi)\, , \Phi^a_0$, where the initial vector $\Phi^a_0$ is arbitrary, to obtain an integral curve and normalize such that the curve parameter $t\in [0;2\pi]$.
\subsection{Approximate Killing Vector Fields}
\label{subsec:akvf}
If the spheroid $S$ is slightly deformed, similar to the initial non-axisymmetric AH in our simulation, no exact solution of (\ref{eq:killing}) exists. But one could try to find a `best match' which minimizes a certain norm of the l.h.s. of (\ref{eq:killing}) on $S$. Such vector fields are often denoted as approximate Killing vector fields (aKVF). Opposed to KVFs there is no unique definition of aKVFs. 
%
Dreyer et al. \cite{Dreyer:2002mx} could show that the Killing transport method is still applicable to yield a `well matching' aKVF. But one has to be aware that the final vector field will not be anymore independent of the particular loops of transportation. Although this effect may be negligible for practical applications, e.g. \cite{Campanelli:2006fy,Rezzolla:2007rd}, if the departure from axisymmetry is `small'. The method has also been used to determine aKVFs in binary black hole initial data, see Caudill et al.~\cite{Caudill:2006coa}.

We found it useful to adapt the coordinate system on the horizon before applying the Killing transport method such that the azimuthal transport revolves the minima of the scalar 2-curvature \footnotemark, see appendix \ref{sec:AdaptedCoordinates}.
\footnotetext{An spheroid has two minima of the scalar 2-curvature which coincide with the minima of the KVF, given by the symmetry axis of the body.}
Another approach to find an approximate Killing vector field has been given by \cite{Cook:2007akv}. They use a variational principle to minimize the `non-symmetric' features of the vector field. A similar method can be found in the appendix of \cite{Lovelace:2008tw}, for an application to a BBH simulation see \cite{Scheel:2008rj}. 
%
%
Recently Beetle \cite{Beetle:2008yt} pointed out that Cook's~\cite{Cook:2007akv} approach is closely related to an older proposal by Matzner~\cite{Matzner:2003ass}, where the aKVF is the solution of an eigenvalue problem. 
An outstanding question is still the normalization of these aKVFs.
An interesting new idea has been given in the appendix of \cite{Lovelace:2008tw}, where the aKVF is normalized to a particular surface integral instead of a single integral curve.

In our approach these difficulties do not appear because no KVF/aKVF is explicitly required to represent the axisymmetry/perturbed axisymmetry. Instead we compute the invariant surface averages $\mu_n$ which exist in any case and from those compute
the IH multipole moments/\emph{generalized} IH multipole moments through the algebraic system linking the two sets of invariants, subsection \ref{subsec:munonaxi}.
\subsection{Coordinate Vector Fields}
\label{subsec:coordinateVFs}
If the coordinates are conveniently adapted to the metric manifold, the coordinate vectors can automatically generate symmetries (if existing), such as the Boyer-Lindquist coordinate vectors $\p_t$ and $\p_\phi$ in a Kerr spacetime. This is also the case 
for the \emph{adapted spherical coordinates} $(\theta_{\rm  asc},\phi_{\rm  asc})$, see appendix \ref{sec:AdaptedCoordinates}, and the particular initial setup we chose in our simulations \footnotemark. Then the coordinate vector field
\begin{equation} \label{eq:phiAC} 
  \Phi^a_{\rm asc} = \p_{\phi_{\rm asc}} \, ,
\end{equation}
is a good approximation to the KVF and we can estimate the spin $J[\Phi^j]\approx J[\Phi^j_{\rm asc}]$ with (\ref{eq:phi}), see the application in section \ref{sec:num}.  
\footnotetext{In general this is not the case and the correct solution of the Killing equation has to be found. In our case coordinate vector fields are very useful for the comparison of section \ref{sec:num}. }

Similarly \cite{Campanelli:2006fy,Caudill:2006coa} use the three rotational Killing vectors of Euclidean space in Cartesian coordinates
\begin{equation} \label{eq:cckvf}
  \Phi^{i[j]}_{\rm cc} = (x^k-C^k) {\epsilon^{ij}}_k \, , \, j=1,2,3 \, ,
\end{equation}
where ${\epsilon^{ij}}_p \delta^{pk}\!=\!\epsilon^{ijk}$ is the flat space Levi-Cevita tensor and $C^j$ a point inside $S$, to define a Euclidean spin vector $(J[\Phi^{i[1]}_{\rm cc}], J[\Phi^{i[2]}_{\rm cc}], J[\Phi^{i[3]}_{\rm cc}] )$ and together with (\ref{eq:phi}) to estimate $J[\Phi^i]\approx J[\Phi^i_{\rm cc}]$, where $J[\Phi^i_{\rm cc}]$ denotes the Euclidean norm of this vector which allows them to study the spin precession in a BBH inspiral and to estimate the final spin after merger.
Referring to \cite{Campanelli:2006fy} this Euclidean spin vector reproduces the Bowen-York spin parameters of the conformally flat initial data and for the final black hole $ \vert J[\Phi^j]-J[\Phi^j_{\rm cc}] \vert \ll 1$ as in our simulations. 

\section{Invariants of the Horizon in Kerr}
\label{sec:Kerr}

Before we go into the details of how the surface averages $\mu_n$ are linked to the IH multipole moments in the next section \ref{sec:multipoles}, we want to remind that the mass $M^{\rm Kerr}_l$ and angular momentum $J^{\rm Kerr}_l$ multipole moments of Kerr $M^{\rm Kerr}_l + i J^{\rm Kerr}_l = m (iJ/m)^l$ are uniquely given by Kerr spin $J$ and mass $m$. In this section we will review the analytic formulas necessary to extract Kerr spin and mass from an AH and give a new formula which we apply in our simulations.

%
In many numerical simulations Kerr spin and mass $(J,m)$ are being computed from the `equatorial' circumference \footnotemark and the area $(L(c_e),A)$ of the BH surface, see \cite{Smarr:1973sgc}. A more recent approach is to use an extremum of the scalar 2-curvature and the area $(\rsc_{\rm ext},A)$, see \cite{Lovelace:2008tw}.
\footnotetext{This is the curve $c_{\rm e}$ along the maximum of $\rsc$ in Kerr.} 
Each of these invariant pairs uniquely determines a Kerr spacetime and is related to the other through the Kerr metric such that we are free to choose the numerically most convenient one.
In order to benefit from exact numerical integration and to avoid interpolation on the horizon we chose the invariants $(\mu_{2} (\rscHat),A)$, see (\ref{eq:sigma}). The explicit algebraic expressions relating $J\, \leftrightarrow\, L(c_{\rm e})\, \leftrightarrow\, \rsc_{\rm  ext}\, \leftrightarrow\, \mu_{2}(\rscHat)\,\,$ ($\leftrightarrow\, \mu_{2}(\ImPsiTwoHat)$) are derived in the following.

Any axisymmetric 2-metric $q_{ab}$ can be put in the compact form
\begin{eqnarray}\label{eq:axi2metric}
 dq^2 = \frac{A}{4\pi} \left( \frac{1}{f(\chi)} d\chi^2 + f(\chi) d\phi^2 \right).
\end{eqnarray}
For the 2-surface of a Kerr black hole $f(\chi)$, see \cite{Smarr:1973sg}, is given by 
\begin{equation} \label{eq:fchi}
 \quad f(\chi) = \frac{1-\chi^2}{1- \hat{\beta}^2 (1-\chi^2)},
 \quad \chi := \cos \theta,
\end{equation}
where $\hat{\beta} \in [0; 1/\sqrt{2}]$ is called the \emph{Kerr distortion parameter}
and $(\theta,\phi)$ are the Boyer-Lindquist spherical coordinates. 
The distortion parameter $\hat{\beta}$ is related to the more familiar dimensionless spin 
parameter $\hat{a}=a/m=J/m^2$ by
\begin{equation}
  \hat{\beta}^2 = \frac{1}{2} \left(1- \sqrt{1-\hat{a}^2} \right) = \frac{\hat{c}^2}{\hat{c}^2 +1 },
\end{equation}
to Kerr spin $J$ and mass $m$ \footnote{For completeness note that $m_{\rm irr}=R_{\rm areal}/2$ is the \emph{irreducible mass} and $R_{\rm areal}=\sqrt{A/(4\pi)}$ the \emph{areal radius}.} by 
\begin{equation} \label{eq:jbeta}
  J = \frac{A}{8\pi}  \sqrt{ \frac{ 1-\sqrt{1-\hat{a}^2 }  }{ 1+\sqrt{1-\hat{a}^2 } } } = \frac{A}{8 \pi} \frac{\hat{\beta}}{\sqrt{1-\hat{\beta}^2}} = : \frac{A}{8\pi}\, \hat{c} \, , \quad 
 m = \frac{1}{2} \sqrt{\frac{A}{4\pi (1-\hat{\beta}^2)}} \, .
\end{equation}
Smarr \cite{Smarr:1973sg} pointed out the analog of the surface of rotating material bodies to the black hole horizon, where the equatorial circumference increases as the body spins up. The equatorial circumference for the Kerr horizon is given by integrating (\ref{eq:axi2metric}) along the maximum of $\rsc$ which is the curve $(\chi=0,\phi)$,
\begin{equation}
  L(c_{\rm e}) = \oint^{2\pi}_0 \sqrt{\frac{A}{4 \pi}\, f(\chi=0)} \,d\phi = \sqrt{\frac{A\pi}{1-\hat{\beta}^2}} = 4 \pi m \, .
\end{equation}
%
For the numerical application in arbitrary coordinates this is practical, if the curve $c_{\rm e}$ is known to overlap with a coordinate line.
If this is not the case the extrema of $\rsc$ are an appealing alternative, see \cite{Lovelace:2008tw,Scheel:2008rj}.
The scalar 2-curvature of $q_{ab}$ (\ref{eq:axi2metric}) is
\begin{equation} \label{eq:rsckerr}
 \rsc = - \frac{8 \pi}{A} \frac{1}{2} f''(\chi) \quad \rightarrow \quad \rscHat = - \frac{1}{2} f''(\chi) \, ,
\end{equation}
with extrema at $\chi_{\rm min}=1;-1\, , \, \chi_{\rm max}=0$. We obtain 
\begin{equation} \label{eq:rsc-extrema}
 \rscHat_{\rm max} = \frac{1}{(1-\hat{\beta}^2)^2}\, , \quad \rscHat_{\rm min} = 1 - 4 \hat{\beta}^2 \, .
\end{equation}
\subsection{An invariant surface integral in Kerr }
\label{subsec:mu2}
If the scalar 2-curvature (or alternatively ${\rm Re} \Psi_2$ , since ${\rm Re} \Psi_2 = -1/4\, \rsc$ for Kerr) has been computed on a finite grid, interpolation is required to obtain the extrema. This is not necessary if the following surface integrals are employed 
\begin{equation} \label{eq:sigmaR} 
 \mur{2}{\rscHat} := \left< \left( \left< \rscHat \right> - \rscHat \right)^2 \right>, 
 \quad \left< \rscHat \right> := \frac{1}{A} \oint_S \rscHat \, dA \, .
\end{equation} 
Moreover, the numerical error of $\mu_2(\rscHat)$ benefits from averaging over all points on the grid and exact numerical integration can be used.
%
With the normalization of (\ref{eq:rsckerr}) the average $<\! \rscHat\! >_{\rm grid} = 1 + \epsilon_{\rm num}$, where $\epsilon_{\rm num}$ is the numerical error, for any 2-metric computed on a finite grid on $S$ according to the Gauss-Bonnet theorem.
For Kerr the integral appearing in (\ref{eq:sigmaR}) is taken over a rational function in $\chi$. We obtain 
\begin{equation} \label{eq:sigma2Kerr}
  \mu_2(\rscHat) = \frac{-15 - 70 \hat{c}^2 +128 \hat{c}^4 + 70 \hat{c}^6 + 15 \hat{c}^8}{80(1+\hat{c}^2)}  +   \frac{3(1+\hat{c}^2 )^4}{16} \frac{\arctan (\hat{c})}{\hat{c} } \, ,
\end{equation}
where $\hat{c}$ is defined in (\ref{eq:jbeta}). In our simulations we compute the surface average $\mu_2(\rscHat)$ numerically and solve (\ref{eq:sigma2Kerr}) for the Kerr $\hat{c}$. Kerr spin and mass are then given by $J=A/(8\pi)\hat{c}$ and $m^2=A(1+\hat{c}^2)/(16\pi)$ (\ref{eq:jbeta}).
For the numerical application in section \ref{sec:num} the Kerr spin deviates significantly from the IH spin during the initial phase but the `non-Kerr' features are radiated during the evolution and finally vanish below the numerical error.


Note that we could similarly use any $\mu_n(\rscHat),\, n>2$ or $\mu_n(\ImPsiTwoHat),\, n>1$ to compute $\hat{c}$ for Kerr. In that case $\ImPsiTwoHat = -\frac{1}{4} g''(\chi),\, g(\chi) := \frac{(1+\hat{c}^2)^2 }{\hat{c}( 1+\hat{c}^2 \chi^2) }$, see \cite{Chandrasekhar83} and some algebra.
It follows that $\mu_1(\ImPsiTwoHat)=0$ and $\mu_1(\chi\cdot\ImPsiTwoHat) =  \hat{c}$. 
The explicit appearance of the Boyer-Lindquist coordinate ($\chi=\cos\theta$) is inconvenient for the numerical application. For $\mu_2(\ImPsiTwoHat)$ we obtain an expression similar to (\ref{eq:sigma2Kerr}) which is $\mu_2(\ImPsiTwoHat)= \frac{-15+170\hat{c}^2+112\hat{c}^4 + 70\hat{c}^6+15\hat{c}^8 }{320 (1+\hat{c}^2) } +\frac{3(1+\hat{c}^2)^4 }{64 } \frac{\arctan(\hat{c})}{\hat{c}}$.
To extract more information than the Kerr $\hat{c}$ we have to consider the whole set of $\mu_n$ and follow the procedure explained in the next section.

\section{Invariants of Axisymmetric Isolated Horizons}
\label{sec:multipoles}

For the calculations in the last section to be reasonable when applied to an AH found in a numerical simulation, we had to assume that the detected 2-surface was in a slice of Kerr.
We relax this condition and allow the spacetime to be dynamical in the vicinity of the horizon which we assume to be an axisymmetric \emph{isolated horizon} (IH) \cite{Ashtekar:2000hw,Ashtekar-Beetle-Lewandowski:2001}.  
On the horizon in Kerr all multipole moments are necessarily given by spin and mass, therefore higher moments contain no extra information. This is in general not the case on an axisymmetric IH, where an infinite set of independent multipole moments permits more complexity, see \cite{Ashtekar:2004gp}.

Ashtekar et al.~\cite{Ashtekar:2004gp} exploit the axisymmetry to define an invariant coordinate system $(\chi, \phi)$ for which the 2-metric has the form (\ref{eq:axi2metric}), $\p_\phi$ is the KVF and the (zonal) harmonics $\{ Y^{l0}(\chi) \}$ represent an orthonormal basis $\oint_S Y^{l0}(\chi) Y^{l'0} (\chi) dA=\frac{A}{4\pi} \delta^{ll'}$ which they use to define the dimensionless IH mass $\hat{I}_l$ and angular momentum $\hat{L}_l$ multipole moments
 \begin{align}
  \label{eq:mmoments}
  & \hat{I}_l  :=  \oint_S 1/4\, \rsc(\chi) Y^{l0}(\chi) \, dA\, , &
  & \hat{L}_l  := - \oint_S \ImPsiTwo(\chi) Y^{l0}(\chi) \, dA \, . \\
  \label{eq:mmdecomposition}
& \rsc(\chi)  = 4\cdot\frac{4\pi}{A} \sum^{\infty}_{l=0} \hat{I}_l \, Y^{l0}(\chi) \, , & 
& \ImPsiTwo(\chi) =-\frac{4\pi}{A} \sum^{\infty}_{l=0} \hat{L}_l \, Y^{l0}(\chi) \, . 
\end{align}
On IHs without matter fields (like in Kerr) the Weyl scalar $\Psi_2$ is invariant and ${\rm Re} \Psi_2 = -1/4\, \rsc$. 

Note that for Kerr $J\cdot 8\pi/A = \hat{c}= \sqrt{1/(3\pi)}\, \hat{L}_1$ and for an IH $J[\Phi^j]\cdot 8\pi/A = \sqrt{1/(3\pi)}\, \hat{L}_1$, where $\Phi^j$ is the KVF corresponding to $(\chi,\phi)$ and $J[\Phi^j]$ given by (\ref{eq:phi}). Therefore, the curvature component $\ImPsiTwo$ is sometimes called \emph{rotational} Weyl scalar and the $\hat{L}_l$ \emph{angular momentum} multipole moments, all vanish in the absence of spin. 

The invariants $\hat{I}_l,\, \hat{L}_l$ are subject to certain algebraic constraints such that $\hat{I}_0=\sqrt{\pi}$ (Gauss-Bonnet), that the mass dipole $\hat{I}_1$ and the angular momentum monopole $\hat{L}_0$ vanish \footnotemark.
If the 2-metric (\ref{eq:axi2metric}) admits a reflection symmetry as for Kerr $f(\chi)=f(-\chi)$, see (\ref{eq:fchi}), all odd $\hat{I}_l$ and even $\hat{L}_l$ vanish, too.
\footnotetext{Therefore, the invariant coordinates are sometimes called `center of mass frame' of the IH.}
%
%
\subsection{The invariants $\mu_n$ on axisymmetric isolated horizons}
\label{subsec:munonaxi}

In analogy to the method explained in subsection \ref{subsec:mu2} for Kerr, where we gave the formula (\ref{eq:sigmaR}) to compute the Kerr $\hat{c}$ from the surface average $\mu_2(\rscHat)$, we would like to relate the invariants $\mu_n(\rscHat),\, \mu_n(\ImPsiTwoHat)$ (\ref{eq:sigma}), which are numerically easy to obtain in any coordinate system, to the IH multipole moments (\ref{eq:mmoments}) which would require the invariant coordinates for a direct computation of the integrals (\ref{eq:mmoments}) (as for example being done in \cite{Vasset:2009pf}). 

%
We obtain the algebraic relations between the  $\mu_n(\rscHat),\, \mu_n(\mathrm{Im } \hat{\Psi}_2)$ and the $\hat{I}_n,\, \hat{L}_n$ by inserting (\ref{eq:mmdecomposition}) into (\ref{eq:sigma}) 
\begin{eqnarray} 
  \label{eq:munI}
  \mur{n}{\rscHat} &=&
  \left< \left( 1 - 2  \sum^\lmaxI_{l=0} \hat{I}_l \, Y^{l0} (\chi) \right )^n \right>
  \, , \quad n=2,3,...,\nmaxI \, , \\
  \label{eq:munL}
  \mur{n}{ \ImPsiTwoHat} &=&
  \left< \left( 0 + \sum^\lmaxL_{l=0} \hat{L}_l \, Y^{l0} (\chi) \right )^n \right>
  \, , \quad n=2,3,...,\nmaxL \, ,
\end{eqnarray}
where we assume that $\rscHat,\, \ImPsiTwoHat$ are given by finite sets of multipole moments up to $\lmaxI,\, \lmaxL$. We obtain \footnote{Here the indices $I\, ,L$ in $\lmaxI,\, \lmaxL,\,\nmaxI,\, \nmaxL  $ are omitted.}
\begin{eqnarray} 
  \mur{n}{\rscHat} &=& 
  \sum^{n}_{m=0} {n \choose m} (-2)^m \sum_{\vert K_\lmax \vert = m} {m \choose K_\lmax} (\hat{I}_{\_})^{K_\lmax} \left< (Y^{\_ 0})^{K_\lmax} \right> \, , \\
 \mur{n}{\ImPsiTwoHat} &=&
 \sum_{\vert K_\lmax \vert = n} {n \choose K_\lmax} (\hat{L}_{\_})^{K_\lmax } \left< (Y^{\_ 0})^{K_\lmax} \right> \, , \quad n=2,3,...,\nmax \, ,
\end{eqnarray}
where $K_{\lmax}=(k_1,k_2,...,k_\lmax)$ is a multi-index of length $\lmax$, ${k \choose k_1,k_2,...}$ is the multinomial coefficient and $(\hat{I}_{\_})^{K_\lmax} \left< (Y^{\_ 0})^{K_\lmax} \right>= (\hat{I}_1)^{k_1} (\hat{I}_2)^{k_2}... \left< (Y^{10})^{k_1} (Y^{20 })^{k_2} ... \right> $ . The integers $\nmaxI\, ,\nmaxL$ match the numbers of non-trivial $\hat{I}_n,\, \hat{L}_n$ given by the algebraic constraints mentioned earlier and $\lmaxI\, , \lmaxL$. 
The coefficients $\left< (Y^{\_ 0})^{K_\lmax} \right>$ are integrals over products of (zonal) spherical harmonics. They are given by the associated Clebsch-Gordan coefficients and higher order generalizations. 

Consider the following example. In a simulation of a perturbed Kerr spacetime we locate the AH and compute the surface integrals $\mu_n(\rscHat)\, , \mu_n(\ImPsiTwoHat)$ (\ref{eq:sigma}) numerically to $\nmax=6$ \footnotemark to equate them with the r.h.s. of (\ref{eq:munI}), where we assume that the 2-surface is a cross-section of an IH with reflection- and axisymmetric 2-metric. Then the algebraic systems (\ref{eq:munI}), (\ref{eq:munL}) become
\footnotetext{Formally the solutions of the algebraic systems depend on $\nmax$. It determines the number of multipole moments we can resolve $\lmax$ and is limited by the numerical noise. In pratice the solution for lower $\lmax$ does not change as we go to higher $\nmax$.}
\begin{eqnarray} \label{eq:example}
  \mur{n}{\rscHat} &=& \left< \left( 1 - 2( \sqrt{\pi} Y^{00} + \sum_{l=2,4,6,8}
  \hat{I}_l Y^{l0} ) + \mathcal{O}_I \right)^n \right>,\, n=2,3,4,5,6 \\
\mur{n}{\ImPsiTwoHat} &=& \left< \left( \sum_{l=1,3} \hat{L}_l Y^{l0} + \mathcal{O}_L
  \right)^n \right>, n=2,4,6\, \, ,
\end{eqnarray}
which we solve for $\hat{I}_2,\,\hat{I}_4,\,\hat{I}_6,\,\hat{I}_8,\,\mathcal{O}_I$ and $\hat{L}_1,\,\hat{L}_3,\,\mathcal{O}_L$,
where $\mathcal{O}_I$, $\mathcal{O}_L$ are constants accounting for the truncation of the expansions.
%
Since we simulate a perturbed Kerr spacetime, we pick the solution that is real and for which $ -\hat{I}_2 > \hat{I}_4 > -\hat{I}_6> \mathcal{O}_I$ and $\hat{L}_1 > -\hat{L}_3 > \mathcal{O}_L$ holds as for Kerr.

%

In analogy to electro dynamics dimensionfull factors can be added to attribute a physical interpretation to the $\hat{I}_l,\, \hat{L}_l$, see \cite{Ashtekar:2004gp}. 
To obtain the spin we need

%
\begin{equation} \label{eq:j1v2}
  J_1 = \sqrt{\frac{1}{12\pi}} \frac{A}{4\pi} \hat{L}_1 \,  . 
\end{equation}
%
The equation $J[\Phi^j]=J_1$ holds if $\Phi^j$ is the KVF corresponding to the invariant coordinates $(\chi,\phi)$.

The surface integrals $\mu_{n}$ are well defined even in the absence of axisymmetric and allow to extend the concept of IH multipole moments by adding the $m\neq0$ harmonics in the expansions of $\rscHat$ and $\ImPsiTwoHat$ on the r.h.s. of \ref{eq:j1v2}. Nevertheless, for the evolution of the non-axisymmetric initial data studied in \ref{sec:num} we assume that the contribution of ode/even mass/angular momentum multipole moments (reflection symmetry), higher harmonics as well as $m\neq0$ harmonics is small and can be accounted for through $\mathcal{O}_I$, $\mathcal{O}_L$. We do not further investigate the possibility of generalized multipole moments.
Our approach aims at numerical convenience and is flexible enough to extract, in principle, other invariants like the generalized multipole moments proposed by Owen~\cite{Owen:2009sb} who considers the eigenfunctions of the intrinsic Laplacian on the horizon. 


\section{Accurate Computation of $\rsc$, $\Psi_2$ on the AH}
\label{sec:computinggeometry}

In this section we will show how to compute the curvature components $\rsc$ and $\Psi_2$ accurately, where we assume that the 3+1 evolution variables \footnotemark extrinsic 3-curvature $K_{ij}$, 3-metric $\gamma_{ij}$ (together with $\p_i K_{jk},\, \p_i \gamma_{jk},\, \p_i \p_j \gamma_{kk'}$) and the horizon coordinate shape $X^j$ are given on a Cartesian grid.
\footnotetext{They can be easily assembled from the BSSN evolution variables.}
The accurate calculation of curvature components on a deformed 2-sphere in a Cauchy slice is a common problem in numerical relativity which appears in horizon finding algorithms. 
%
Various methods have been tried to discretize the necessary spatial derivatives $\p_jh$, $\p_i\p_jh$ by finite differencing, finite element, pseudo-spectral and spectral methods, using squared $(\theta,\phi)$ grids or multipatch grids, for a review see
\cite{thornburg}.
Our approach is motivated by the work of \cite{Baumgarte:1996iah}. 
There a spectral decomposition of the coordinate shape function $h(\theta,\phi)$ is being used to compute Cartesian derivatives. The 1st derivatives $\p_jh$ are necessary to obtain a surface triad $\{ s^i,\, u^j,\, v^k \}$ (required to compute the Weyl scalars) and the 2nd derivatives $\p_i\p_jh$ to obtain the extrinsic 2-curvature $^2\!K_{ij}$ of $S$ embedded into the Cauchy slice (additionally required to compute the scalar 2-curvature).

If we parametrize the AH with spherical coordinates, the embedding $X^j(\theta,\phi)$ into the Cartesian grid is 
\begin{equation} \label{eq:embeddingX}
 X^j(\theta,\phi) = h(\theta,\phi)\, n^j + C^j \, ,
\end{equation}
where $C^j$ is a coordinate location inside the horizon (for example the coordinate centroid), $n^j$ the Cartesian radial unit vector $n^j=\frac{1}{r} x^j, \,\, r=\sqrt{\delta_{ij}x^i x^j}$ and $x^j$ are Cartesian coordinates. 

\subsection{Spectral decomposition}
To compute spatial derivatives one could decompose $h(\theta,\phi)$ into
\begin{equation} \label{eq:hylm}
 h(\theta,\phi) = \sum^{\lmax}_{l=0} \sum^{-l}_{m=l} [h]^{lm} Y^{lm}(\theta,\phi) \, , 
\end{equation}
where $[h]^{lm}$ are the expansion coefficients and $Y^{lm}$ the spherical harmonics.
The evaluation of $\p_j Y^{lm}(\theta,\phi)$ would require the Jacobian to transform between spherical and Cartesian coordinates. This is inconvenient for the numerical application, since the Jacobian is singular at the spherical coordinate poles.

Therefore, \cite{Baumgarte:1996iah} take a tensor basis which is build of the radial unit vector $n^i(x^j)= x^j/r$ and thus defined in Cartesian coordinates (and easily parametrized with any other local coordinate system on the 2-surface, e.g. spherical $n^i(\theta,\phi)=(\sin\theta \cos\phi, \sin\theta \sin\phi,\cos\theta )$ or stereographic coordinates $n^j(u,v)=(2u, 2v, u^2+v^2-1 )/(1+u^2+v^2 ) $),
\begin{equation} \label{eq:nkl}
 h = \sum^{l_{max}}_{l=0} [h]^{K_l} N_{K_l} \, ,
\end{equation}
%
where $K_l$ is again a multi-index of length $l$, $N_{K_l}= n_{k_1} n_{k_2} \ldots n_{k_l}$ is the vector product of unit vectors and the location-independent coefficients $[h]^{K_l}$ are symmetric tracefree tensors (STF), the notation is adapted from \cite{Thorn:1980}.
%
If the STFs are known, they can be translated to obtain the expansion (\ref{eq:nkl}), for how to $[h]^{lm} \leftrightarrow [h]^{K_l} $ see \cite{Baumgarte:1996iah}. 
The partial derivative of the tensor product $\p_j N_{K_l}$ consists of the derivatives  $\p_i n_j= (\delta_{ij} - n_i n_j)/r$. In detail the implementation of the STF tensors and its derivatives is a bit cumbersomely but straight forward.

We use another basis of the harmonics instead $( \delta_{ij} n^i \Nlm^j )^l$, where $\Nlm^j$ is a constant complex Euclidean null vector $(\Nlm_j \Nlm^j)=0,\, \Nlm^j \neq 0$, see Sec.11.5.1., Vol.II \cite{Batemanmanuscriptproject:1953htf} or \cite{lachiezerey:2004nbe}.
The expression $( n_j \Nlm^j )^l$ is a homogeneous harmonic polynomial of Euclidean space of order $l$, therefore  $\Delta_{\rm flat} ( n_j \Nlm^j )^l = 0$. The radial vector $n^j$ defines a restriction of the polynomial to the unit sphere $x^ix^j\delta_{ij}=1$. 
It is known that such restrictions are eigenfunctions of the Laplacian of the induced metric (this applies to any embedding of $S^2$ into Euclidean space, e.g. an ellipsoid). On the unit sphere this implies $\Delta_\circ( n_j \Nlm^j )^l= l(l+1)( n_j \Nlm^j )^l$, where $\Delta_\circ$ is the Laplacian of the standard spherical 2-metric.
This holds for any null vector $\Nlm^j$.
In order to span each $l$-eigenspace of $\Delta_\circ$ with $2l+1$ linear independent eigenfunctions we define a list of null vectors  
\begin{equation}
  \Nlm^j_{[lm]} = \left( i \sin(m a_l), i \cos (m a_l), 1 \right), \quad
  a_l = \frac{2\pi }{2l+1} ,\quad  m = -l,\cdots,l \,\, ,
\end{equation}
where the roots of unity have been used such that the $\Nlm^j_{[lm]},\,$ have the Euclidean norm $\Nlm_{j} \Nlm^{j} = -| e^{i \frac{2\pi m}{2l+1} } |^2 + 1$. Now we can define the new basis $\Phi^{lm}:= ( n_j \Nlm^j_{[lm]} )^l$ and decompose $h$ into
\begin{equation} \label{eq:nlm} 
  h = \sum^{\lmax}_{l=0} \sum^{-l}_{m=l} [h]^{lm}_\Nlm \, ( n_j \Nlm^j_{[lm]} )^l \, .
\end{equation}
The $\Phi^{lm},\, m=-l,\cdots,l$ are not orthogonal in each $l$-eigenspace but across different eigenspaces. They are related to the standard basis by
\begin{eqnarray} \label{eq:ylmphilm}
 Y^{lm} &=& B^{l m} \sum^l_{m' = -l} \Phi^{l m'} e^{-i\, m' m \,  a_l}, \\
 \Phi^{lm} &=& \frac{1}{2l+1} \sum^l_{m'=-l} \frac{Y^{lm'}}{B^{lm'}} e^{i\, m' m\,  a_l}, \\ \notag 
 B^{l m} &=& (-1)^{m} \frac{1}{l!} \sqrt{\frac{(l+m)!(l-m)!}{4\pi (2l+1)}}
 \, . \notag
\end{eqnarray}
and we can transform the coefficients $[h]^{lm} \leftrightarrow [h]^{lm}_\Nlm$. Derivatives of the new basis \footnotemark are given by
\begin{eqnarray} 
 \p_k \Phi^{lm} &=& ( n_j \Nlm^j )^{l-1} l\, ( \p_k n_j \Nlm^j ) \\
 \p_k \Phi^{lm} &=& ( n_j \Nlm^j )^{l-1} l\,  \frac{1}{r} \left(\Nlm_k - n_k n_j\Nlm^j \right) \,,
\end{eqnarray}
and similarly for higher derivatives $\p_i\p_j\Phi^{lm}$.
\footnotetext{Here we omit the subscripts $\Nlm^j_{[lm]} \rightarrow \Nlm^j$ for simplicity.}
\subsection{Surface triad}

Now we have the Cartesian derivatives $\p_j h,\, \p_j\p_i h$ at hand and are able to compute the outward pointing surface normal $s^j=\gamma^{jk} s_k$ 
\begin{equation} \label{eq:triad}
 s_j = \lambda ( n_j - \p_j h ),\, \lambda = 1/\sqrt{ \gamma^{ij} ( n_i - \p_i h ) ( n_j - \p_j h ) } \, .
\end{equation}
%
In order to complete the surface triad $\{s^i,u^j,v^k\}$ we set $u^j= \frac{1}{\sqrt{\gamma_{ik}\p_\theta X^i \p_\theta X^k}} \p_\theta X^j$ and $v^k= \varepsilon^{ijk} s_i u_j$, where $\varepsilon^{ijk}=\|\gamma \|^{-1/2} [123]^{ijk}$ is the spatial Levi-Civita tensor and $[123]^{ijk}$ the pure
alternating symbol.
\subsection{Extrinsic and intrinsic 2-Curvature}

The extrinsic 2-curvature $^2\! K_{ij}$ of $S$ embedded into the Cauchy slice is given by
\begin{equation}
 ^2\!K_{ij} = D_i s_j - s_i s^k D_k s_j \, ,
\end{equation}
where the second derivatives $\p_j\p_k h$ are required and the Christoffel symbols associated with the 3-metric to compute the 3-covariant derivative $D_j$.
Then the intrinsic 2-curvature $\rsc$ is given by Gauss' \emph{theorema egregium}
\begin{equation} \label{eq:gausstheorem}
  \rsc = \mathcal{R} -2 R_{ij} s^i s^j + {^2\!\mathcal{K}}^2 - {^2\!K^{ij}}\, {^2\!K_{ij}} \, , 
\end{equation}
%
where $^2\!\mathcal{K} = {^2\!K_{ij}} q^{ij}$ and $q^{ij} = \gamma^{ij} - s^i s^j$ is the induced 2-metric in Cartesian components (also required to raise the indexes of $^2\! K_{ij}$ in the last summand on the r.h.s. of (\ref{eq:gausstheorem})) and $R_{ij}$, $\mathcal{R}$ are the 3-dimensional Ricci tensor and scalar.
\subsection{Area Element}

The computation of surface integrals on the AH requires the area element $dA=\sqrt{\det q_{ab}}\,  d\theta d\phi$, where we need the induced 2-metric in local coordinates
\begin{equation}
 q_{ab} = \p_a X^j \p_b X^k \gamma_{jk} \, ,
\end{equation}
here $X^j$ has been defined in (\ref{eq:embeddingX}), for an alternative see appendix of \cite{Baumgarte:1996iah}.
\subsection{$\Psi_2$ and other Weyl scalars}
\label{sec:Weyl}

To obtain mass and angular momentum multipoles (\ref{eq:mmoments}) an accurate computation of $\Psi_2$, given the 3+1 evolution variables, is requried. Additionally, we want to follow the constraints $\Psi_0=0$ and $\Psi_1=0$ which hold for Kerr and on IHs \cite{Ashtekar:2000hw} in the simulation of section \ref{sec:num}.
The \emph{electric} $E_{ij}$ and \emph{magnetic} $B_{ij}$ parts of the Weyl tensor $C_{ijkl}$ w.r.t. time-like normal $\tilde{n}^\mu$ of the Cauchy slice are
\begin{eqnarray}
  E_{ij} &\equiv& - C_{ijkl} \tilde{n}^k \tilde{n}^l 
  = -R_{ij} + {K_i}^k K_{kj} - \mathcal{K}\,  K_{ij} \, ,\\
  B_{ij} &\equiv& -\star C_{ijkl} \tilde{n}^k \tilde{n}^l 
  = - {\varepsilon_i}^{kl} D_k K_{lj} \, .
\end{eqnarray}
We further project $E_{ij},\,B_{ij}$ onto the surface triad $\{ s^i,u^j, v^k\}$ and obtain the Weyl scalars, see \cite{kelly:mpn,smarr:1975sgr},
\begin{eqnarray}
 \Psi_2 &=& -\frac{1}{2} (E_{jk} - i B_{jk}) s^j s^k \, ,\\
 \Psi_0 &=& - (E_{jk} - i B_{jk}) m^j m^k \, ,\\
 \Psi_1 &=& -\frac{1}{\sqrt{2}} (E_{jk} - i B_{jk}) m^j s^k,
\end{eqnarray}
where $m^j= \frac{1}{\sqrt{2}} ( u^j - i v^j) $.

We monitor the dynamics of the AH during the evolution in section \ref{sec:num} by computing the dimensionless surface integrals
\begin{equation} \label{eq:psin}
  \hat{\psi}_0 = \oint_S \arrowvert \Psi_0 \arrowvert \, dA,\quad
  \hat{\psi}_1 = \oint_S \arrowvert \Psi_1 \arrowvert \, dA,\quad
  \hat{\psi}_2 = \left\arrowvert \frac{1}{8\pi} \oint_S 4 {\rm Re}\,\Psi_2 \,dA + 1 \right\arrowvert \, ,
\end{equation}
which vanish for a MOTS in a slice of Kerr or an IH.

\section{Numerical Evolution and Initial Data}
\label{sec:num}

In order to test and compare the new techniques we applied them to the dynamical AH of a non-axisymmetric spinning BH in a 3+1 simulation ringing down to Kerr which as been carried out using the CCATIE code \cite{Pollney:2007ss}. This is a 3D finite differencing code based on the Cactus Computational Toolkit \cite{allen2000ccp}. 
%
The CCATIE code provides a collection of modules (\emph{thorns}) which allow us to use \emph{puncture initial data} \cite{Brandt:1997tf} with the TwoPunctures thorn \cite{Ansorg:2004ds}, to do time evolution using the BSSN evolution system \cite{Nakamura87,Shibata95,Baumgarte99}, to set proper gauge conditions (where we used 1+log slicing and a hyperbolic gamma-driver condition stemming from \cite{Alcubierre:2002kk} but with advection terms \cite{Pollney:2007ss}), to successively refine the Cartesian mesh with several nested static boxes around the AH (where we used the Carpet AMR driver \cite{carpet-author-and-web-site}) and to locate the horizon every few time steps during the evolution \cite{Thornburg95}.
The horizon finding thorn provides the shape function $h(\theta,\phi)$ which is being used by a separate thorn to interpolate (4th-order Lagrange) all necessary 3+1 evolution variables onto the spherical grid, to accurately compute the curvature components $\rsc,\, \ImPsiTwo$ at the horizon (see section \ref{sec:computinggeometry}) and, finally, to determine the associated quasi-local IH multipole moments using the surface integrals $\mu_n$ (\ref{eq:sigma}).
%


%

\subsection{Initial Data and Grid Parameters}

In order to model the common horizon after the coalescence of an arbitrarily aligned BBH system we chose as a non-trivial initial configuration a misaligned spinning puncture with a nearby smaller non-spinning companion puncture, where the common horizon is already present on the initial slice. 
The Bowen-York parameters of the first puncture are $m_{1}= 0.8 M,\, | s_{1}|= 0.3 M^2$ with orientation $(\theta_{s_1}\! =\! 0.6,\, \phi_{s_1}\! =\! 0.4)$ in the Cartesian grid. And for the second puncture we set $m_2=0.2 M,\, s_2 =0$.

The evolution is being carried out using the method of lines with 4th-order Runge-Kutta time integrator and 4th-order centered stencils for spatial differentiation with the Cartesian grid resolutions $\Delta x = 0.048M,\, 0.035M,\, 0.025,\, 0.02M$ (finest AMR resolutions).
To determine the KVF/aKVF we use the Killing transport method \ref{subsec:Ktransport} with 2nd-order centered stencils for differentiation and a 2nd-order Runge-Kutta integrator, see \cite{Dreyer:2002mx} for more details. 
To compute spatial derivatives of the shape function $h(\theta,\phi)$ we use its decomposition into Spherical Harmonics where the spectral resolution is fixed to $\lmax=10$.
To compute the surface averages $\mu_n$ we use an exact integration scheme, see appendix \ref{sec:NumericalIntegration}, and fix $\nmax=6$.
%
For every Cartesian resolution we use three different spherical horizon grid resolutions $N_\theta \times N_\phi = N_S = 480,\,1104,\, 4900$, where $N_S$ is the total number of grid points on the surface and $N_\phi = 2 (N_\theta+1)$. The horizon finder is using a projective 6-patch grid \cite{Thornburg95} with approximately the same number of points as on the spherical grid.

\subsection{Numerical Evolution}
\subsubsection{Monitoring the Isolation Constraints}
\label{subsec:psinEvolution}
\begin{figure}[t]

\includegraphics[width=0.5\textwidth]{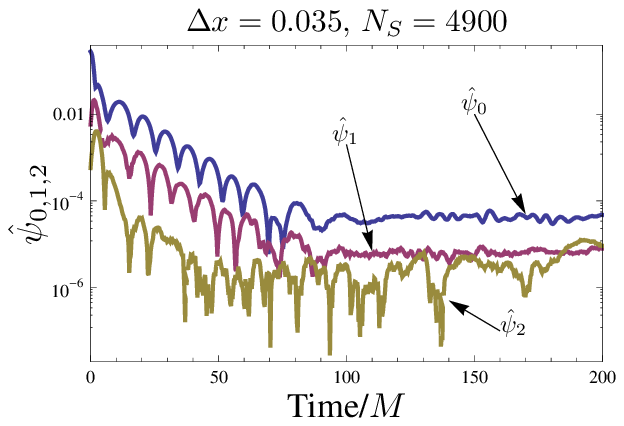}~\includegraphics[width=0.5\textwidth]{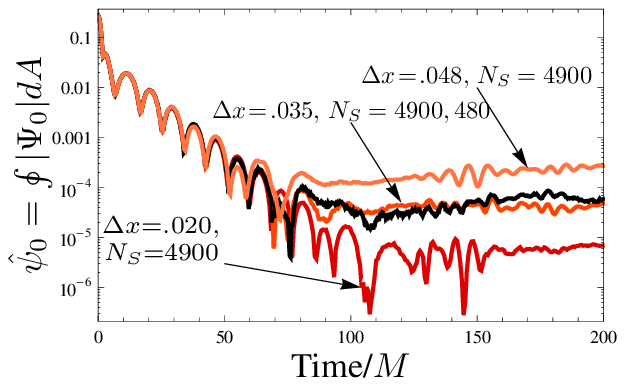}

\caption{Left: time evolution of dimensionless surface averaged Weyl scalars $\hat{\psi}_{0,1,2}$, Right: time evolution of $\hat{\psi}_0$ for 3 different Cartesian resolutions}
\label{fig:psin}
\end{figure}

To monitor the dynamics on the horizon we computed the surface integrals (\ref{eq:psin}) shown in figure \ref{fig:psin} (for Kerr $\hat{\psi}_{0,1,2}=0$).
On the left we see the typical exponentially damped oscillation of the \emph{radiative} Weyl scalars $\Psi_0,\,\Psi_1$ which are (after an initial burst $\hat{\psi}_{0,1} \ll 1$) given by a superposition of several quasinormal-modes, predominately $l=2$ modes, that have been excited by the specific initial data. 
As a fit to the ring-down profile of $\hat{\psi}_0$ we obtain the frequency $\omega_{\rm fit}\approx 0.355+0.088 i$, in agreement with the $l=2$-mode frequencies $\omega_{l=2mn}$, see \cite{Leaver:1985ax}, which are $\omega_{2-20} \approx 0.34+0.089 i,\, \omega_{220} \approx 0.36+0.089 i,\, \cdots$ for the case $J=0.3,\, m=1.035$. 
After around $t>90M$ the perturbations are to weak to be further resolved limited by the total numerical error, which we downsize by increasing the Cartesian grid resolution, see figure \ref{fig:psin} on the right, in order to see the dynamics below $\hat{\psi}_0<10^{-5}$.
For $\Delta x= 0.035$ (black and orange) we computed $\hat{\psi}_0$ for two different spherical resolutions to show that the total error of $\hat{\psi}_0$ (and similar for surface integrals of other curvature components) is almost independent of the spherical resolution due to the spectral methods involved.

%
%
%
\subsubsection{Evolution and Convergence of the Invariants $\mu_n$}

\begin{figure}[t]

  \includegraphics[width=0.5\textwidth]{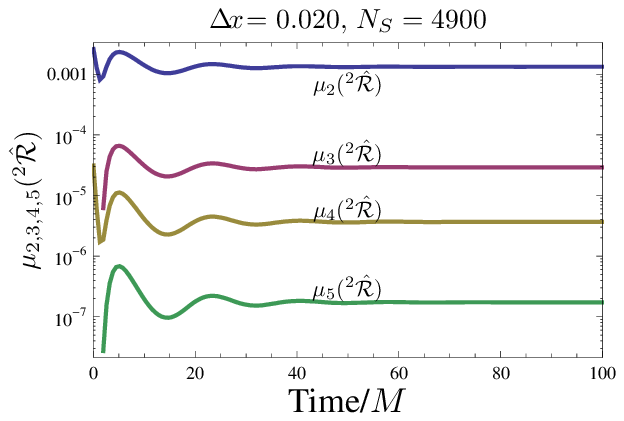}~\includegraphics[width=0.5\textwidth]{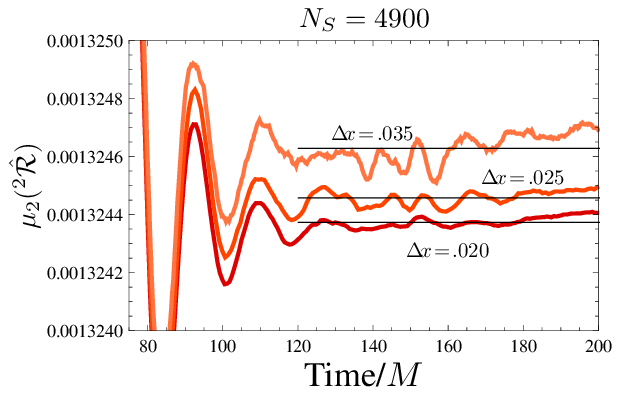}

\caption{Left: time evolution of the surface averages $\mu_{2,3,4,5}(\rscHat)$ over powers of the scalar curvature of the horizon,
        Right: time evolution of $\mu_{2}(\rscHat)$ for 3 different Cartesian resolutions and time averages (straight lines) between $120M-200M$ for each resolution}

\label{fig:mun}
\end{figure}

In figure \ref{fig:mun} we see the exponentially damped oscillation of the $\mu_n$ as they ring-down to their final Kerr value. On the right it is shown how the time averages of $\mu_2(\rscHat) $ (120$M$-200$M$, straight black lines) converge with the expected 4th-order ($4.01$) as the Cartesian grid resolution increases after the oscillations have settled down. 
Apparently, the error of $\mu_{2}(\rscHat)$ does not converge uniformly but the effect flattens out as the Cartesian resolution increases. 
\subsubsection{Evolution of Mass and Angular Momentum Multipole Moments}
\begin{figure}

  \includegraphics[width=0.5\textwidth]{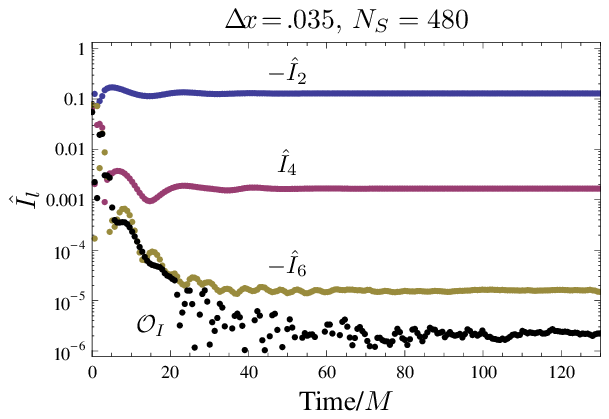}~\includegraphics[width=0.5\textwidth]{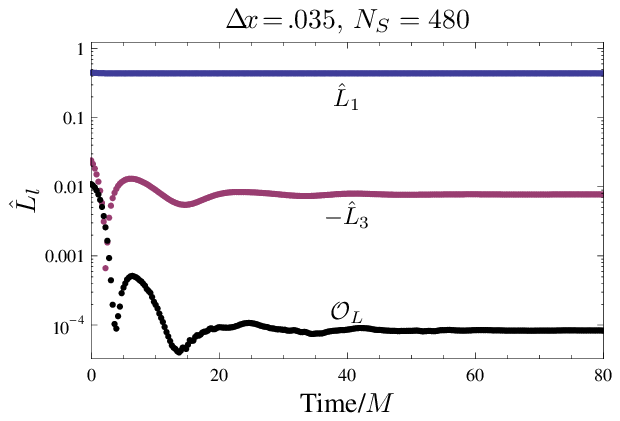}

  \caption{Left: time evolution of mass $\hat{I}_l$, Right: and angular momentum multipole moments $\hat{L}_l$ given as a solution of the algebraic system (\ref{eq:example}) for the $\mu_n$ up to $\nmax=6$; $\mathcal{O}_I$, $\mathcal{O}_L$ account for all higher multipole moments}
\label{fig:mms}
\end{figure}
From the $\mu_n$ we compute the IH multipole moments $\hat{I}_l,\, \hat{L}_l$ corresponding to an reflection and axisymmetric horizon by solving the algebraic system (\ref{eq:example}), where $\mathcal{O}_I$, $\mathcal{O}_L$ account for all higher, non-axisymmetric and non-reflection symmetric multipole moments.
%
%
It is apparent in figure \ref{fig:mms} that these multipole moments are quickly radiated $t<30M$, leaving the horizon almost reflection and axisymmetric but still oscillating. 
%
%
Interestingly, the dimensionless IH spin $\hat{L}_1$ is almost constant during the evolution, as the horizon area (not plotted, $A\approx4\pi\cdot 2.05^2 M^2$). 

\subsubsection{Spin Evolution and Comparison with other methods}
\begin{figure}

\includegraphics[width=0.5\textwidth]{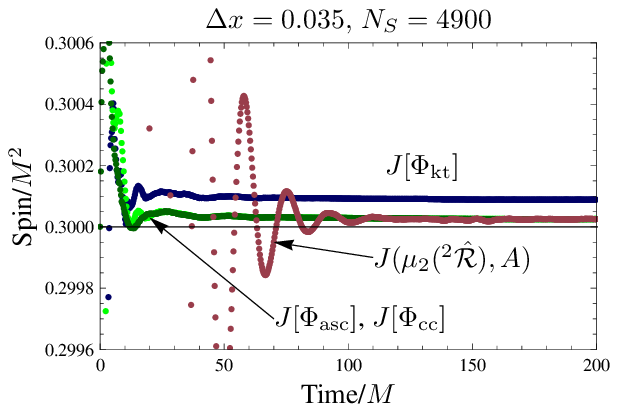}~\includegraphics[width=0.5\textwidth]{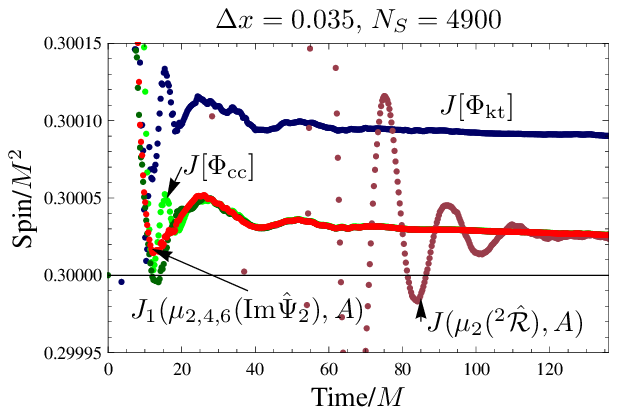} \\
\includegraphics[width=0.5\textwidth]{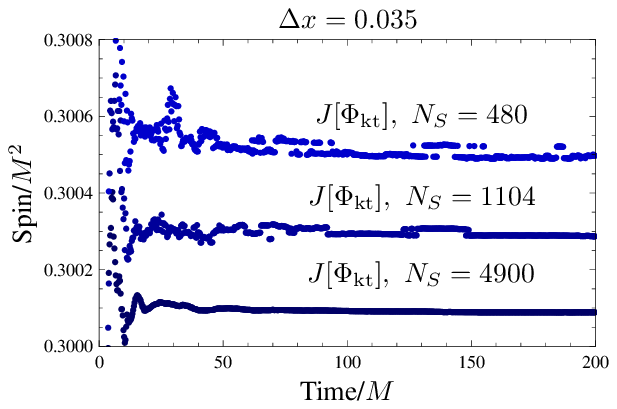}~\includegraphics[width=0.5\textwidth]{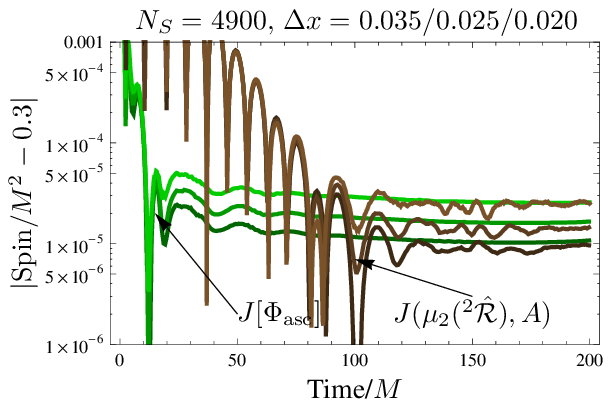}

\caption{
Top Left: time evolution of spins given by the Killing transport aKVF $\Phi_{\rm kt}$, the coordinate vector fields $\Phi_{\rm asc},\, \Phi_{\rm cc}$ and the Kerr spin computed from $\mu_2(\rscHat)$;
Top Right: Zoom of `Top Left' together with angular momentum dipole $J_1 = A/\sqrt{192 \pi^3} \hat{L}_1 $ (red) computed from $\mu_2,\,\mu_4,\,\mu_6$ of $\ImPsiTwoHat$;
Bottom Left: convergence of $J[\Phi_{\rm kt}]$ varying number of spherical grid points $N_S$;
Bottom Right: convergence of $J[\Phi_{\rm asc}]$, $J(\mu_{2}(\rscHat),A)$ varying Cartesian resolution $\Delta x$}
\label{fig:spins}
\end{figure}

In figure \ref{fig:spins} we see the comparison between the various spin measures and their convergence. We have
\begin{enumerate}
  \item $J_1=A/\sqrt{192\pi^3} \hat{L}_1$ (red) computed from the $\mu_n$, (\ref{eq:example}), assuming an axisymmetric IH,
  \item $J[\Phi_{\rm kt}]$ (blue) computed from the Killing transport KVF/aKVF $\Phi_{\rm kt}$, (\ref{eq:transport}),
    (\ref{eq:phi}), assuming an axisymmetric IH,
  \item $J[\Phi_{\rm cc}]$ (light green), $J[\Phi_{\rm asc}]$ (dark green) given by the coordinate vector fields $\Phi_{\rm cc}$
    (Cartesian coordinates), (\ref{eq:cckvf}), $\Phi_{\rm asc}$ (adapted spherical coordinates), (\ref{eq:phiAC}),
    assuming `small' coordinate distortions,
  \item $J=J(\mu_{2}(\rscHat),A)$ (brown) computed from $\mu_2(\rscHat)$, (\ref{eq:sigmaR}), assuming a Kerr horizon.
\end{enumerate}
After a short initial bust all methods yield nearly the same spin value, which stays constant during the evolution; except $J(\mu_{2}(\rscHat),A)$ (brown) which oscillates with the quasinormal frequency. During this phase the horizon seems to be best modelled assuming an axisymmetric dynamical horizon but not Kerr.
%
We chose the numerical setup such that the coordinate distortions are small and $J[\Phi_{\rm asc}]$, $J[\Phi_{\rm cc}]$ overlap with the invariant measure $J_1$. This is in general not the case in a full BBH simulation and these methods should be used with care.

In figure \ref{fig:spins} (bottom right) we see the expected 4th-order convergence (w.r.t. Cartesian grid) of $J[\Phi_{\rm asc}]$, $J[\Phi_{\rm cc}]$ and $J(\mu_{2}(\rscHat),A)$ towards $0.3 M^2$. The convergence of $J_1$ is not shown explicitly. It is a smooth function of the $\mu_n$ (convergence shown above) and converges therefore at the same rate.
%
On the other hand $J[\Phi_{\rm kt}]$ converges at 2nd-order (w.r.t. the spherical grid), figure \ref{fig:spins} (bottom left) \footnotemark, because the Killing transport method requires finite differencing on the horizon grid to determine $\Phi^j_{\rm kt}$. 
\footnotetext{Note that the low resolution $N_S = 480$ (light blue) is to coarse to be in the convergence regime.}
%
%
\section{Conclusion}
\label{sec:Conclusion}

The dominant part of the gravitational radiation at Scri is contained in the quadrupole moment of $\Psi_4$ which is in practice extracted at `large' coordinate spheres around the source in numerical simulations.
Similarly, the dipole moment of the rotational Weyl scalar $\ImPsiTwo$ encodes the quasi-local angular momentum measured at the apparent horizon in the presents of axisymmetry.
The local coordinates on the horizon are in general distorted and a solution of the Killing equation is required to determine an invariant coordinates system in which the multipole moments can be computed.

It is involved to determine the Killing vector field, in particular, to find a convenient approximant in case the axisymmetry is perturbed.
We have shown a new method to extract the horizon multipole moments using coordinate invariant surface integrals $\mu_n$ from which we deduce the multipole moments as a solution of an algebraic system.
%
In case of an axisymmetric IH the angular momentum dipole $J_1$ is equal to the spin $J[\Phi]$ given by a solution of the Killing equation $\Phi^j$ in agreement with our simulations.
Interestingly, the spin of the aKVF $\Phi_{\rm kt}$ (given by the Killing transport method) and the angular momentum dipole moment $J_1(\mu_n(\ImPsiTwoHat),A)$ (given by the $\mu_n$) agree even in the absence of axisymmetry.



There seems to be a dynamical phase of the horizon in which it is better modelled by an axisymmetric dynamical horizon and not with Kerr. Nevertheless, after the horizon is settled the Kerr formula is valid. Then the computation of the Kerr spin using the surface average $\mu_2(\rscHat)$ (or $\mu_2(\ImPsiTwoHat)$) is sensible and numerically more convenient than using the horizon circumference.
%
%
The deviations from Kerr oscillate in agreement with black hole perturbation theory, until they are no more resolvable due to numerical errors. Then the dipole moment of the rotational Weyl scalar agrees with the Kerr spin and the $\mu_n$ take their final Kerr values.

We have shown how to use spectral methods, in a 3+1 finite differencing code, to accurately compute curvature components at the horizon and to extract spin and other multipole moments saving computational costs.
%
%
These techniques, in particular, the non-standard basis of spherical harmonics and the exact integration scheme, should be considered for wave extraction on coordinate spheres or constant mean curvature spheres \cite{Schnetter:2003pv,Metzger:2004pr}. 
%
%

\acknowledgments 

I am pleased to acknowledge Badri Krishnan for his advice and fruitful ideas. I am grateful to Luciano Rezzolla, Jos\'e Luis Jaramillo, Denis Pollney, Geoffrey Lovelace, Frank Ohme, Sascha Husa and Emma Robinson for helpful comments and discussions. I thank Erik Schnetter for providing the IsolatedHorizon thorn as a basis for my implementations. I also thank the anonymous referees for carefully reading the manuscript. Computations were performed at the LRZ Munich using the CCATIE code principally developed at the numerical relativity group at the AEI Potsdam. This work was supported by the Max Planck Society.

\appendix

\section{Exact Integration Schemes for Spherical Grids}
\label{sec:NumericalIntegration}

It is well know that the equation
\begin{equation}\label{eq:gauss}
 \int^b_a f(x) w(x) dx = \sum^N_{i=1} w_i f(x_i) \, ,
\end{equation}
holds exactly, where $w(x)$ is called the \emph{weight function}, if $f(x)$ is a polynomial of degree less than $2N$ and the weights $w_i$ and abscissas $x_i$ are chosen in accordance with the orthogonal basis of polynomials on $[a,b]$ defined by the scalar product $< f|g> := \int^b_a f(x) g(x) w(x) \,dx$, because there are $2N$ degrees of freedom to make both sides of (\ref{eq:gauss}) match, see for example \cite{Szegoe:1939}.

For the integration with $w(x)=1$ on the circle $a=b$, the `correct' weights and abscissas are particularly simple. They are $N$ equi-distant points with equal weights.
This can not be generalized for the integration on the 2-sphere
\begin{equation} \label{eq:fxy}
  \oint_{S^2} f(x,y)\, dA = \sum^{N_S}_{i=1} w_i f(x_i,y_i) \, ,
 \end{equation}
for arbitrary $N_S$, because the number of uniform grid structures is finite $N_S=4,6,8,12,20$, corresponding to the faces of the platonic solids. Since this is a 2D integration, we have $3N_S$ degrees of freedom in the sum on the r.h.s. of (\ref{eq:fxy}) and $(\lmax+1)^2$ spherical harmonics of degree $\le \lmax$. This means if $f(x,y)$ was given by an expansion up to $\lmax$, we needed at least $N_S = (\lmax+1)^2/3$ points to make (\ref{eq:fxy}) hold. Lets say $f(x)$ was given by an expansion of $(7+1)^2-4$ spherical harmonics, then the integration (\ref{eq:fxy}) on an icosahedral grid $N_S=20$ with equal weights would be exact.
There is an extensive body of work on the problem of optimal integration schemes for 
$N_S>20$ (\emph{cubature problem}), see for example \cite{Sobolev:1974}.

There are less optimal compromises available, which require much more points than $(\lmax+1)^2/3$, but which are defined on regular spherical $(\theta,\phi)$ grids.
For example the Gauss-Legendre/Gauss scheme, where the integration along each interval $[-1,1],\, [0;2\pi]$ is a Gaussian quadrature 
\begin{equation} \label{eq:int}
 \oint^{2\pi}_0 \int^{1}_{-1} f(\chi,\phi) d\chi d\phi =
 \sum^{N_{\chi}}_{i=1} \sum^{N_{\phi}}_{j=1} w^{\chi}_i w^{\phi}_j f(\chi_i, \phi_j) \, ,
\end{equation}
where again $\chi=\cos\theta$, $N_S = N_\theta \times N_\phi$ and $N_\phi=2 N_\theta$.

As before the $\phi$-integration is a Gaussian quadrature for $\phi_j = 2\pi (j-1)/N_\phi,\, j=1,...,N_{\phi}$ and equal weights $w^{\phi}_j = 2\pi/N_\phi$, the $\chi$-integration (in that case called Gauss-Legendre quadrature) for $\chi_i$ being the roots of the Legendre polynomials (according to the weight function $w(\chi)=1$). The corresponding weights $w^\chi_i$ can be found in e.g. \cite{press1990nrf}. This method is exact for polynomials of degree less than $2N_\theta$ (less than $\sqrt{2 N_S}<\sqrt{3 N_S}$).

An alternative integration scheme has been found by \cite{Driscoll:1994cft}\footnotemark. There the integration grid is a standard equi-angular $(\theta,\phi)$ grid, $\theta_j= (j-1/2)\pi/N_\theta$ (staggered) and the computation of the roots of the Legendre polynomials not necessary. The weights for even/odd $N_\theta$ are given by
\footnotetext{The authors make use of the fact that the points $\chi_j=\cos\theta_j$ (although not the zeros of the Legendre polynomials on $[1;-1]$) are the zeros of the Chebyshev polynomials of the 1st kind.}
\begin{eqnarray}
 w^{\theta}_j &=& 4/N_{\theta} \sum^{N_{\theta}/2-1}_{k=0} \frac{1}{2l+1}\sin\left( (2k+1) \theta_j \right)\, ,  \quad N_\theta \, {\rm even,} \\
 w^{\theta}_j &=& 4/N_\theta \left( \frac{1}{2 N_\theta} \sin(N_\theta \cdot \theta_j) + \sum^{(N_\theta-1)/2-1}_{k=0}
 \frac{1}{2l+1}\sin\left( (2k+1)\theta_j \right) \right)\, ,\quad N_\theta \, {\rm odd,}
\end{eqnarray}
which allows for exact integration of harmonics of order less than $N_{\theta}/2$ (less than ${\sqrt{1/8 N_S}<\sqrt{2N_S}<\sqrt{3N_S}}$). 
Then equation (\ref{eq:int}) becomes
\begin{equation}
 \oint^{2\pi}_0 \int^{\pi}_{0} f(\theta,\phi) \sin\theta\,  d\theta\, d\phi = 
 \sum^{N_{\theta}}_{i=1} \sum^{N_{\phi}}_{j=1} w^{\theta}_i w^{\phi}_j f(\theta_i, \phi_j) \sin\theta_j.
\end{equation}

A small summarizing example: for the total of $N_S=512$, $N_\theta \times N_\phi = 16 \times 32$ the cubature limit is at $39\approx \sqrt{3\cdot 512} = \sqrt{3 N_S}$, for the Gauss/Gauss-Legendre scheme we get $\lmax<32 = 2 N_\theta$ and for the scheme of \cite{Driscoll:1994cft} we have $\lmax<8 = N_\theta/2$ (we get almost 
the same limit on an icosahedral grid \footnotemark with only $N_S=20$, where $\lmax< 8\approx \sqrt{3\cdot20}$).
%
\footnotetext{Therefore, if one is only interested in the first coefficients of a smooth function on the sphere up to $\lmax=6$, an icosahedral grid with equal weights would be a good choice.}

\section{Adapted Spherical Coordinates}
\label{sec:AdaptedCoordinates}
\begin{figure}[t]
 \includegraphics[width=0.33\textwidth]{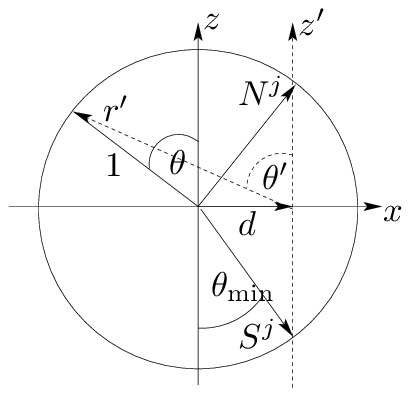}~\includegraphics[width=0.33\textwidth]{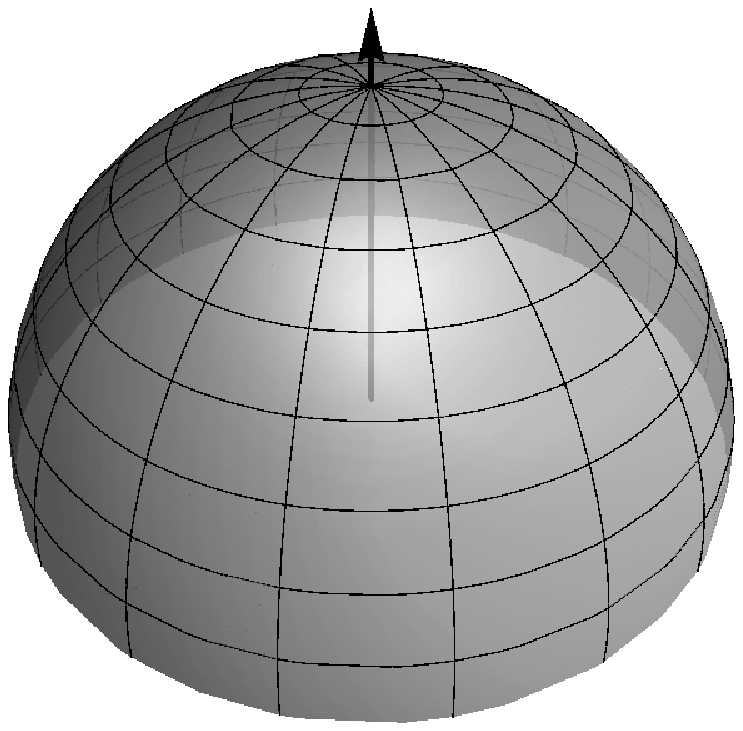}~\includegraphics[width=0.33\textwidth]{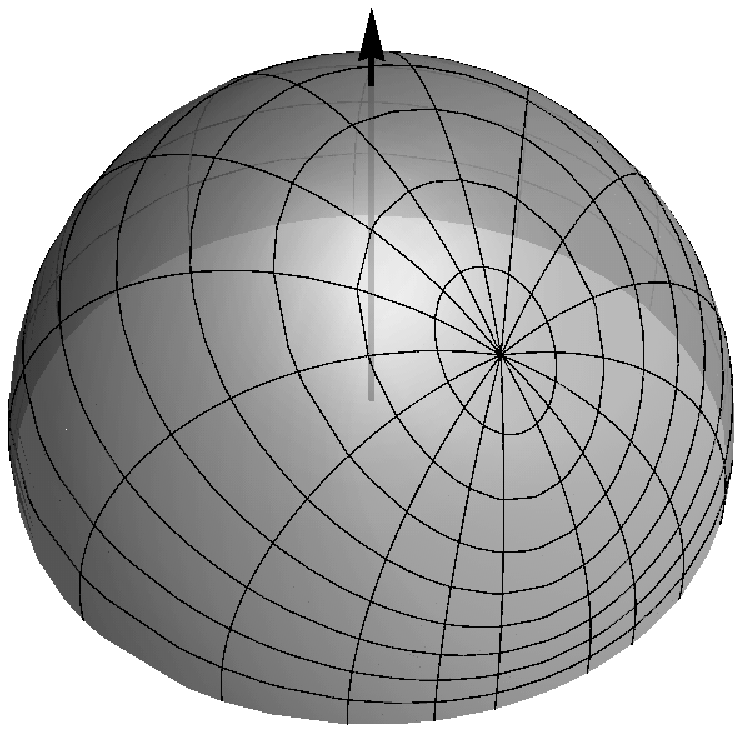}
 \caption{parametrization of the unit-sphere with a shifted spherical coordinate system}
 \label{fig:asc}
\end{figure}

Before solving the 2D Killing equation on a sphere it is useful to have the 2-metric or the horizon shape in a convenient coordinate representation, which is `roughly' adapted to the axisymmetry. Such that the poles of the spherical coordinates system agree with the two minima of the scalar 2-curvature. We assume $\rsc(\theta,\phi)$ to be given on a spherical coordinate system $(\theta,\phi)$, where the two minima are already in the $xz$-plane symmetric to the $x$-axis at $N^j=(\sin\theta_{\rm min},0,\cos\theta_{\rm min})$ and $S^j=(\sin\theta_{\rm min},0,-\cos\theta_{\rm min})$, see figure \ref{fig:asc}. This can always be accomplished by a simple Euler rotation. 
 
In order to obtain the adapted spherical coordinates system $(\theta',\phi')$, we have to shift the Cartesian $z$-axis along the $x$-axis by the amount $d:=\sin\theta_{\rm min}$.
This is being done by
\begin{equation} \label{eq:nndash}
 n_j(\theta,\phi) = r'(\theta',\phi') \, n_j'(\theta',\phi') + d \cdot(1,0,0) \, ,
\end{equation}
where $n_j(\theta,\phi)= (\cos \phi \sin \theta, \sin\phi \sin\theta, \cos\theta),\, 
n_j'(\theta',\phi')= (\cos \phi' \sin \theta', \sin\phi' \sin\theta', \cos\theta') $
are the radial unit vectors in the corresponding coordinate system.
The distance $r'(\theta',\phi')$ is given by
\begin{equation}
 r'(\theta',\phi') = \sqrt{d^2_\pp - 2 r_\pp d_\pp \sin\theta + r^2_\pp } \, ,
\end{equation}
where $d_\pp$, $r_\pp$ are given by
\begin{eqnarray}
r_\pp &=& \cos\phi' \cos \phi + | \sin\phi' | \sqrt{1-\cos^2\phi}\, , \\
d_\pp &=& d \cos\phi' \, .
\end{eqnarray}
And finally, $\cos\phi$ and $\sin\theta$ in terms of $\theta',\phi'$ are given by
\begin{eqnarray}
\cos\phi &=& d \sin^2 \phi' + \cos \phi' \sqrt{1-d^2 \sin^2\phi' } \, , \\
\sin \theta &=& \frac{1}{r_\pp} \left( d_\pp \cos^2\theta' + \sin\theta' \sqrt{r^2_\pp - d^2 \cos^2 \theta'} \,\right)  \, .
\end{eqnarray}
The inverse transformation is given by interchanging $\theta\leftrightarrow\theta'$, $\phi\leftrightarrow\phi'$ $d\leftrightarrow -d$ in the above expressions.


\bibliographystyle{apsrev}

\bibliography{main}

\end{document}